\documentclass[aps,prl,floats,twocolumn,showpacs,superscriptaddress]{revtex4-1}   
\usepackage[dvips]{graphicx}
\usepackage{epsfig}
\usepackage{subfigure}
\usepackage{amsmath}
\usepackage{amssymb}
\usepackage{epsfig}
\usepackage{color}

\begin{document}

\title{Dissipation driven quantum phase transitions and symmetry breaking}
\author{Julia Hannukainen}
\author{Jonas Larson}
\affiliation{Department of Physics, Stockholm University, Se-106 91
  Stockholm, Sweden} 
  \date{\today}

\begin{abstract}
By considering a solvable driven-dissipative quantum model, we demonstrate that continuous second order phase transitions in dissipative systems may occur without an accompanying spontaneous symmetry breaking. As such, the underlying mechanism for this type of transition is qualitatively different from that of continuous equilibrium phase transitions. In our model, the transition is solely a result of the interplay between Hamiltonian and dissipative dynamics and manifests as a non-analyticity in the steady state $\hat\rho_\mathrm{ss}$ in the thermodynamic limit. Expectations of local observables are continuous but typically with discontinuous first order derivatives in agreement with second order phase transitions. While the model, a large number of driven two-level systems under collective dissipation, is in some sense fully connected, mean-field results are incapable of capturing the full picture. 
\end{abstract}
\maketitle

{\it Introduction.} -- The concept of spontaneous symmetry breaking~\cite{RG} plays a central role in physics, ranging from classical thermodynamics to the standard model of high energy physics. In the Landau theory of equilibrium phase transitions (PTs)~\cite{LT}, the symmetry broken phase is characterized by a non-zero local order parameter, while the unordered phase is identified by a vanishing order parameter. This mean-field theory also predicts that the physics in the vicinity of the critical points is entirely described by a few critical exponents. Scale invariance and the renormalization group provide additional understanding of critical behaviour and especially its universal properties -- continuous PTs (second order transitions) can be grouped into different university classes where their properties depend only on macroscopic properties, i.e. dimensionality and symmetries~\cite{RG}. The importance of dimension and symmetry is also manifested in the Mermin-Wagner theorem~\cite{AA}. For long, the conventional wisdom was that the above ideas formed a complete description of continuous equilibrium PTs, but with the discovery of topological PTs and Kosterlitz-Thouless transitions it was understood that not all continuous PTs are accompanied with a symmetry breaking nor a local order parameter~\cite{RG,topological}.   

Recently one type of non-equilibrium PTs that occur in driven-dissipative quantum systems has especially gained much attention due to its relevance to well controlled quantum optical experiments~\cite{openPT1,openPT3,ion}. Here, the non-analyticity, characteristic of the PT, appears in the system's non-equilibrium steady state (NESS) $\hat\rho_\mathrm{ss}$ rather than its ground state as for quantum PTs~\cite{sachdev}. By tailoring the system-environment couplings and the driving, it is possible to prepare a desirable $\hat\rho_\mathrm{ss}$~\cite{ion} and hence, also causing behaviours reminiscent of continuous PTs~\cite{openPT1}. Naturally, compared to equilibrium PTs very little is known about this new non-equilibrium quantum critical behaviour. For a system Hamiltonian $\hat H$ supporting a quantum PT (QPT), the environment implies additional fluctuations that may prohibit the build-up of long range order according to the fluctuation-dissipation theorem and thereby forbids continuous PTs~\cite{lro}. Here the dissipation acts as an effective temperature, and according to the Mermin-Wagner theorem a PT in lower dimensions can be ruled out, at least for the breaking of a continuous symmetry. Nevertheless, criticality may survive due to the fact that the extensive entanglement is relatively robust to noise at the critical point~\cite{ehud}. Even when criticality survives, the universality, i.e. critical exponents, may be altered by the environment~\cite{critDicke2}, as may properties of the phases~\cite{openPT2}. When the critical behaviour of the NESS can be connected to the QPT of the system Hamiltonian, it seems rather general that the NESS undergoes a similar symmetry breaking at the critical point as the ground state of $\hat H$~\cite{sbness}. Hence, a local order parameter can be defined. However, the lack of a Noether's theorem for driven-open systems~\cite{opensym} suggests that the whole idea of symmetries must be handled differently for NESS PTs. The situation becomes more delicate when the PT itself results from the coupling to the environment, i.e. it stems from the interplay between unitary and dissipative dynamics~\cite{openPT2,openPT1b,openPT1c}. In this scenario the state $\hat\rho_\mathrm{ss}$ can typically not be linked to phases of the Hamiltonian and their properties may be very distinct from equilibrium states. It then comes natural to ask whether the same type of mechanisms behind equilibrium PTs, e.g. symmetry breaking, apply also to dissipation driven NESS PTs. A first step along these lines was taken in~\cite{openPT1c} by examining possible critical exponents. As for~\cite{openPT1b}, in Ref.~\cite{openPT1c} a one dimensional free fermionic gas was considered such that criticality can only arise due to the dissipation and it was found that the correlation length diverges with a critical exponent $\lambda=1/(K-1)$ with $K$ the the number of sites upon which local dissipation acts. A rather general discussion for symmetry breaking for dissipative quantum systems was given in Ref.~\cite{sbness}, and it was argued that as long as the model supports a certain symmetry a breaking is expected. The present Letter demonstrates, however, that continuous PTs may indeed occur without any symmetry breaking in driven-dissipative systems. As such, these transitions fall outside the Landau paradigm of PTs. 

{\it Dissipation driven PTs.} -- To date, engineered driven-dissipative systems are mainly found in the AMO community, especially trapped ions~\cite{ion} and cold atoms~\cite{atom}. These experiments are well described by a Markovian Lindblad master equation~\cite{bruer}
\begin{equation}\label{master}
\frac{\partial}{\partial t}\hat\rho\!=\!\hat{\mathcal{L}}\left[\hat\rho\right]\!\equiv\!i\!\left[\hat\rho,\hat H\right]+\sum_i\gamma_i\left(2\hat L_i\hat\rho\hat L_i^\dagger-\hat L_i^\dagger\hat L_i\hat\rho-\hat\rho\hat L_i^\dagger\hat L_i\right)\!,
\end{equation} 
where we have defined the Liouvllian $\hat{\mathcal{L}}$. The first term on the right represents the unitary evolution generated by $\hat H$ (system Hamiltonian plus Lamb shifts), while the second term incorporates the effects stemming from the coupling to the environment with the decay rates $\gamma_i$ ($\geq0$) and `jump' operators $\hat L_i$. 

Equilibrium QPTs can be traced to a non-analyticity of the ground state for some critical coupling $g_c$ in the thermodynamic limit. For a dissipative system the ground state is replaced by the steady state $\hat\rho_\mathrm{ss}$ of Eq.~(\ref{master}), and a PT is again marked by a non-analyticity emerging in the thermodynamic limit. As for the standard equilibrium classification of phase transitions, for a continuous (2'nd order) NESS PTs the expectation values of local observables $O=\mathrm{Tr}[\hat{\mathcal O}\hat\rho_\mathrm{ss}]$ should be continuos but with possible discontinuous first order derivatives.  

If the Liouvillian has no zero eigenvalues, the unique steady state is the maximally mixed one $\hat\rho_\mathrm{ss}=\mathbb{I}/D$ with $D$ the Hilbert space dimension. This state is also clearly a steady state for any Hermitian jump operators $\hat L_i=\hat L_i^\dagger$. If further $[\hat H,\hat L_i]=[\hat L_i,\hat L_j]=[\hat L_i,\hat L_j^\dagger]=0$ $\forall i,\,j$ the steady states are diagonal in the energy eigenbasis, i.e. $\langle\varepsilon_n|\hat\rho_\mathrm{ss}|\varepsilon_m\rangle=p_n\delta_{nm}$ for some weights $p_n$ and with $|\varepsilon_n\rangle$ the $n$'th eigenstate of $\hat H$. In particular, the ground state $|\varepsilon_0\rangle$ is a `dark state' that is transparent to the effect of the environment. Such a model describes dephasing in the energy basis, and criticality does not derive from environmental fluctuations.  Alternative to the above, criticality driven by the environment stems from non-cummutability among $\hat H$ and the jump operators. In this scenario, $\hat\rho_\mathrm{ss}$ is not necessarily a simultaneous dark state of the jump operators and an eigenstate of the Hamiltonian. Regardless of the situation, the existence and especially the uniqueness of steady states of Lindblad master equations are relevant questions that have been explored in the past~\cite{ss}. It is only recently, however, that general properties of $\hat\rho_\mathrm{ss}$ in terms of PTs and novel phases of matter have been explored.

{\it Model system.} -- The ideas of scale invariance and local order parameters rely on local Hamiltonians, e.g. tight-binding models. At the critical point the characteristic length diverges as $\xi^{-1}\sim|g-g_c|^\lambda$ and the energy gap closes as $\Delta\sim|g-g_c|^{z\lambda}$ for the dynamical critical exponent $z$~\cite{sachdev}. For `fully connected models', i.e. all particles are connected, locality is in a strict sense lost. Nevertheless, it is still possible to show scale invariance, and also to introduce a counterpart of $\xi$ that has been termed `coherence number'~\cite{fully}. In addition, we can still talk about local observables provided that it can be expressed as a sum $\hat{\mathcal O}=\sum_i\hat o_i$ where $\hat o_i$ is restricted to act on particle $i$. For a continuous PT we thereby require that all local $\hat{\mathcal O}$'s are continuous. 

Here we consider a model that was frequently discussed in the late 70's in terms of cooperative emission of radiation and how this relates to optical bistability~\cite{model1,model1b}. The model is analytically solvable in the sense that the (unique) steady state is obtainable~\cite{model2}. It comprises $K$ driven two-level atoms/qubits collectively coupled to a leaky photon mode, and after elimination of the photon degrees of freedom one derives the Lindblad master equation~\cite{model3,model4}  
\begin{equation}\label{systemmaster}
\frac{\partial\hat\rho}{\partial t}=i\omega\!\left[\hat\rho,\hat S_x\right]+\frac{\kappa}{S}\left(2\hat S_-\hat\rho\hat S_+-\hat S_+\hat S_-\hat\rho-\hat\rho\hat S_+\hat S_+\right).
\end{equation}
The $S$-operators are the collective spin operators $\hat S_\alpha=\sum_{i=1}^K\hat\sigma_\alpha^{(i)}$ with $\hat\sigma_\alpha^{(i)}$ the $\alpha=x,\,y,\,z$ Pauli matrix acting on qubit $i$ and the collective raising/lowering operators $\hat S_\pm=\hat S_x\pm i\hat S_y$. Furthermore, $\omega$ is the drive frequency, $\kappa$ the effective fluorescence decay rate, and $S=K/2$ the total spin. The Hamiltonian alone, $\hat H=\omega\hat S_x$, is trivial, and likewise is the Lindblad part of Eq.~(\ref{systemmaster}) on its own. Energetically the Hamiltonian supports the state $|S,-S\rangle_x$ (the spins pointing down along the $x$-direction), while the dissipation pushes the state towards $|S,-S\rangle_z$ (the spins pointing down along the $z$-direction). Any PT between these limiting states is a result of the interplay between the unitary and dissipative dynamics -- note especially that $[\hat S_x,\hat S_-]\neq0$. Whether a non-analyticity emerges as $S\rightarrow\infty$ is {\it a priori} not clear. Indeed, the model Hamiltonian $\hat H_\mathrm{mod}=\hat H_0+g\hat H_1=\hat S_x+g\hat S_z$, that shares the same limiting ground states as our open model, is clearly not critical. 

The total spin is preserved for Eq.~(\ref{systemmaster}), and its unique steady state can be expressed in terms of the spin operators as~\cite{model2}
\begin{equation}\label{steads}
\hat\rho_\mathrm{ss}=\hat\eta\hat\eta^\dagger,
\end{equation}
where $\hat\eta=\frac{1}{\sqrt{D}}\sum_{n=0}^{2S}\left(\frac{\hat S_-}{g^*}\right)^n$, $g=i\omega S/\kappa$, and the normalization
\begin{equation}
D=\sum_{m=0}^{2S}\frac{(2S+m+1)!(m!)^2}{(2S-m)!(2m+1)!}|g|^{-2m}.
\end{equation}  
We may note that the model possesses the dual symmetry $\omega\rightarrow-\omega$ and $\hat S_-\leftrightarrow\hat S_+$ with corresponding steady state as (\ref{steads}) with the raising/lowering operators interchanged. This observation may seem trivial and irrelevant, but it will be reflected in the analysis below when considering the mean-field solutions.

{\it Mean-field and finite size solutions.} -- The thermodynamic limit $S\rightarrow\infty$ is usually associated with the classical limit of the spin and we thereby expect mean-field methods to correctly predict the critical exponents. Hence, quantum fluctuations alone cannot cause the destabilization of the different phases, which is indeed not uncommon for fully connected models~\cite{jonaselinor}.  

In the simplest mean-field picture, quantum correlations are fully discarded and we treat operators as commuting quantities. By normal ordering the equations, this is equivalent to assigning a coherent state ansatz of the state. The resulting equations-of-motion follow from $\partial_tO\equiv\partial_t\langle\hat{\mathcal O}\rangle=\mathrm{Tr}\left[\hat{\mathcal O}\partial_t\hat\rho\right]$ for any operator $\hat{\mathcal O}$, e.g. for the spin variables
\begin{equation}\label{eom}
\begin{array}{l}
\displaystyle{\frac{\partial S_x}{\partial t}=2\frac{\kappa}{S} S_xS_z},\\ \\
\displaystyle{\frac{\partial S_y}{\partial t}=-\omega S_z+2\frac{\kappa}{S}S_yS_z},\\ \\
\displaystyle{\frac{\partial S_z}{\partial t}=\omega S_y-2\frac{\kappa}{S}\left(S_x^2+S_y^2\right)}.
\end{array}
\end{equation}
The steady state solutions (fixed points) for $\lambda\equiv\omega/2\kappa\leq1$, taking into account that the total spin is preserved, are
\begin{equation}\label{mfss1}
\left(S_x,S_y,S_z\right)=S\left(0,\lambda,\pm\sqrt{1-\lambda^2}\right),
\end{equation}
with the solution $S_z=-\sqrt{1-\lambda^2}$ being stable and with the other unstable~\cite{sm}. In the parameter regime $\lambda\geq1$ the steady states are
\begin{equation}\label{mfss2}
\left(S_x,S_y,S_z\right)=S\left(\pm\sqrt{1-1/\lambda^2},1/\lambda,0\right).
\end{equation}
Here, however, the solutions are not stable. Neither of the above bifurcations agree with the more familiar ones, e.g. a pitchfork bifurcation in which a single stable solution turns into two stable and one unstable or a Hopf where a stable solution becomes two periodic solutions with complex conjugated eigenvalues~\cite{strogatz}. The linear stability analysis shows that the branches $S_x=\pm\sqrt{1-1/\lambda^2}$ have purely imaginary eigenvalues, but this is still not representing a Hopf bifurcation in the sense that the solutions do not approach limit cycles. At this level of mean-field study, the absence of a stable steady state for $\lambda>1$ is clearly in contrast to the full quantum solution~(\ref{steads}). Similar observations have been found for optical bistability where on a mean-field level the solutions form a saddle-node bifurcation, while the full quantum solution does not show the typical hysteresis behavior~\cite{quantbist}. Walls and co-workers suggested that for $\lambda\geq1$ it is more relevant to consider the time averaged solutions in order to define physical observables~\cite{model1}. 
The steady state solutions~(\ref{mfss1}) and~(\ref{mfss2}) and their stabilities are depicted in Fig.~\ref{fig1} (a).     

It is clear that the critical point occurs for $\lambda_c=1$, and that the critical exponent for the magnetization $S_z$ $\delta=1/2$. By linearizing around the stable solution for $\lambda\leq1$ it is also possible to extract the mean-field dynamical critical exponent~\cite{sm}. In particular, taking into account the preserved spin the eigenvalues of the corresponding Jacobian are $\lambda_{1,2}=-2\kappa\sqrt{1-\lambda}$, which give the characteristic time scale $T\propto(1-\lambda)^{-1/2}$.  

Already the mean-field results signal that the criticality of the model falls outside the Landau paradigm for continuous PTs. The magnetization (or polarization) $S_z$ serves as our order parameter and we call the phase for $\lambda<1$ `magnitized', while the other phase will be termed `incoherent' for reasons to become clear. It is found that $S_z$ obeys the typical behaviour for a continuous PT; it has a discontinuous first derivative and close to the critical point it is determined by a critical exponent. In fact, any local observable will be continuous. The same holds true for the critical slowing down result which is characterized by a diverging time scale $T$ at $\lambda=1$. However, the stability properties of the steady state solutions imply that there is no apparent symmetry that could be spontaneously broken~\cite{sm}. In contrast to Hamiltonian systems, for an irreversible master equation such as the Lindblad one, a symmetry does not automatically define a preserved quantity. More precisely, a symmetry for a Lindblad master equation is defined as invariance of the Liouvillian $\hat{\mathcal{L}}$ under some unitary $\hat U$, i.e. $\hat{\mathcal{L}}$ is not altered by $\hat H\rightarrow\hat U\hat H\hat U$ and $\hat L_i\rightarrow\hat U\hat L_i\hat U$~\cite{opensym}.  The steady state solutions (\ref{mfss1}) and (\ref{mfss2}) are symmetric under a $\pi$ rotation around $S_y$, $S_x\leftrightarrow -S_x$ and $S_z\leftrightarrow -S_z$. This is exactly the dual symmetry mentioned above, but it is evident that this cannot represent a true symmetry since the two solutions have different stability properties. Thus, our model lacks symmetries.

\begin{figure}
\centerline{\includegraphics[width=8cm]{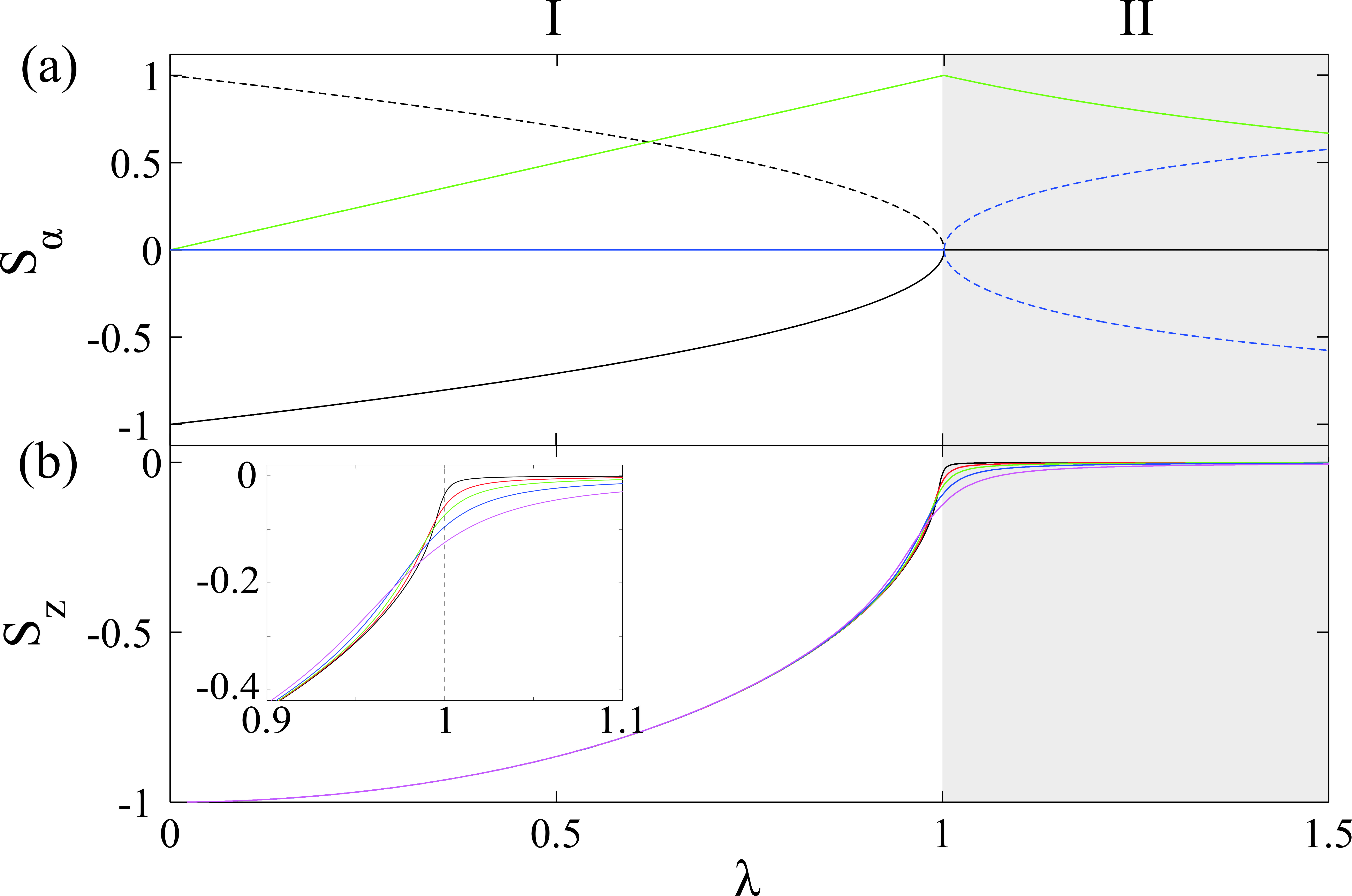}}
\caption{(Color online)  Mean-field (a) and full quantum steady state solutions (b) as a function of $\lambda$. For the mean-field expectations, $S_x$, $S_y$, and $S_z$ are marked by the colours blue, green, and black respectively, and furthermore the solid lines represent stable solutions while the dashed ones unstable. The two phases, magnetized and incoherent, are marked by I and II (gray shaded). In (b) the different curves give the magnetization $S_z$ for different spins: $S=50$ (magenta), $S=100$ (blue), $S=200$ (green), $S=400$ (red), and $S=1600$ (black). The inset shows a close up of $S_z$ in the vicinity of the critical coupling.} 
\label{fig1}
\end{figure}

Even though the mean-field analysis indicates a new type of critical behavior, it might fail to correctly describe the full quantum problem. The question is whether the results carry over to the full quantum problem. Indeed, as we have seen the mean-field predictions do not reproduce the correct result for $\lambda>1$ also in the thermodynamic limit. We thereby turn to numerics and compare and contrast the mean-field results to the quantum ones for large spin values. 

The magnetization $S_z$ is displayed in Fig.~\ref{fig1} (b) for different values of $S$. For $\lambda<1$ the full quantum solutions converge towards the mean-field one in the thermodynamic limit. A scaling analysis also confirms that the critical exponent derived from the full quantum solution $\delta=1/2$~\cite{sm}. In the regime where the Hamiltonian dominates, $\lambda>1$, the solution in the thermodynamic limit becomes $\left(S_x,S_y,S_z\right)=S\left(0,1/\lambda,0\right)$, which is clearly different from the mean-field result. The variances $\Delta S_\alpha=\left(\langle\hat S_\alpha^2\rangle-\langle\hat S_\alpha\rangle^2\right)/S^2$ scale as $S^{-1}$ for $\lambda<1$ and approaches $1/3$ asymptotically for large $\lambda$~\cite{sm}. Thus, when $\kappa\ll\omega$ the steady state becomes maximally mixed and thereby the name incoherent phase for $\lambda>1$. At first this is counterintuitive since the leading term is the Hamiltonian. However, $\hat\rho_\mathrm{ss}$ is the state at $t\rightarrow\infty$ regardless of how fast or slow it approaches the steady state, and hence also a weak decohering mechanism can have substantial influence at long times. The various spin variances do not scale equally, and furthermore the corresponding exponents do not attain simple fractions which one could expect for a fully connected model~\cite{sm}.

{\it Quantum correlations.} -- Another universal feature of QPTs is how entanglement properties scale close to the critical point~\cite{entcrit}. In particular, the entanglement is maximized at the critical point, both for short and infinite range models. It has been shown that for fully connected models, where mean-field predictions gives the correct exponents, non-trivial quantum correlations may exist away from the critical point also in the thermodynamic limit~\cite{vidal}. Criticality in the present model derives from large fluctuations from its intrinsic open character and one could therefore expect them to completely demolish any quantum correlations in the thermodynamic limit. 

By fully characterizing the entanglement of a multipartite state we would need to partition it in all possible ways and calculate the corresponding entanglement between its constituents. Here we focus on qubit-qubit entanglement measured by the negativity $\mathcal N$~\cite{neg}. Negativity is both a necessary and sufficient condition to quantify entanglement for pairwise mixed qubit states. As a symmetric state, $\mathcal N$ is numerically easy to calculate~\cite{molmer,sm}. The result for the scaled negativity $N=\mathcal N/\mathrm{max}({\mathcal{N}})$ is shown in Fig.~\ref{fig2} for different system sizes. Away from the critical point ($\lambda<1$), the negativity scales as $\mathcal{N}\sim S^{-1}$ with the system size, while for the maximum $\mathrm{max}(\mathcal{N})\sim S^{-0.9}$ such that it actually vanishes identically in the thermodynamic limit. The inverse scaling $\sim S^{-1}$ demonstrates the phenomenon of `shared entanglement' which states that a large entanglement cannot be obtained among all constituents simulataneously~\cite{eshare}.  As we verify in~\cite{sm}, by calculating the spin squeezing, which serves as an entanglement witness for the full state, the `total' amount of entanglement contained in $\hat\rho_\mathrm{ss}$ is approximately constant with $S$ away from the critical point. Both the negativity and the squeezing peak at the critical point~\cite{model3,model4}, which is expected from the general behavior known from continuous QPTs~\cite{entcrit}.

\begin{figure}
\centerline{\includegraphics[width=8cm]{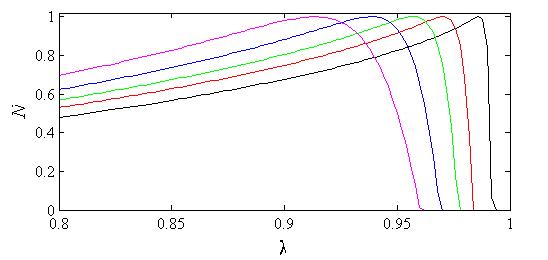}}
\caption{(Color online) Normalized negativity $N=\mathcal N/\mathrm{max}({\mathcal{N}})$ as a measure of qubit-qubit entanglement for different spin numbers $S$. The color marking for the different curves are the same as Fig.~\ref{fig1}. As $S\rightarrow\infty$ the entanglement peaks at the critical point $\lambda=1$. For smaller values of $\lambda$ the negativity goes asymptotically to zero.} 
\label{fig2}
\end{figure}

{\it Conclusion.} -- In this work we demonstrated that a new type of continuous PTs can appear in driven-dissipative quantum systems. In particular, the system shows evidence of a continuous PT in terms of emergent non-analytic behaviors in the thermodynamic limit in both the system state and local observables. Still the transition cannot be tied to a symmetry breaking as in the Landau theory of continuous PTs. Absence of symmetry breaking in continuous PTs has previously been discussed in terms of topological PTs where a local order parameter cannot be assigned to the system~\cite{topological}. Here, we do introduce a `local' order parameter in the magnetization, but it should not be taken as an order parameter in the strict sense since it cannot be associated to a symmetry in the first place. Recently the physics behind symmetry breaking in dissipation driven PTs has been explored~\cite{sbness}. Why the present model does not belong to their rather general results is because it does not support a symmetry to start with, i.e. one can observe continuous NESS PTs even in systems lacking symmetries. In classical systems, non-equilibrium PTs may take place without breaking of any symmetry, but these transitions are typically dynamical, which means that the system evolution can display non-analytic behavior upon changing some parameter~\cite{classpt}. As a next step, our results should be tested on more general grounds, i.e. not for a fully connected model but a local one, for example those of Refs.~\cite{openPT1b,openPT1c}. It is believed that the same type of behavior will be recovered also in these models. 

The physical realization of our model was discussed in detail in~\cite{model3}, and with the present experiments with cold atomic condensates loaded in optical resonators~\cite{eth} or Raman coupled cold atomic gases~\cite{raman} it should indeed be possible to reach the critical point with these setups. Both setups have the advantage that the internal energy between the atomic states can be tuned to zero. The magnetization $S_z$ is also directly measurable via either time-of-flight detection or fluorescence detection in the two respective experiments. Measuring other local observables follow directly from applying the desirable pulses first~\cite{haroche}.

\begin{acknowledgements}
We thank Gerard Milburn, Thomas Kvorning, and Chitanya Joshi for helpful discussions. We acknowledge financial support from the Knut and Alice Wallenberg foundation (KAW) and the Swedish research council (VR).  
\end{acknowledgements}

\pagebreak
\setcounter{equation}{0}
\setcounter{figure}{0}
\setcounter{page}{1}
\makeatletter

\begin{center}
\large
{\bf SUPPLEMENTARY INFORMATION}
\normalsize
\end{center}

\section{Mean-field stability analysis}
The stability of the steady state solutions, Eqs.~(6) and (7) of the main text, is given by linearizing around these solutions and explore the eigenvalues of the corresponding Jacobians~\cite{strogatzsm}. Since the total spin is preserved it is convenient to turn to the canonical variables $(z,\phi)=(\cos\theta,\phi)$, for which the equations-of-motion become
\begin{equation}
\begin{array}{l}
\displaystyle{\frac{\partial z}{\partial t}=-2\kappa\left(1-z^2\right) +\omega\sqrt{1-z^2}\sin\phi,}\\ \\ 
\displaystyle{\frac{\partial\phi}{\partial t}=-\omega\frac{z}{\sqrt{1-z^2}}\cos\phi.}
\end{array}
\end{equation}
$\theta$ and $\phi$ are the polar and azimuthal angels, and note in particular that $z$ is identical to the magnetization $S_z$. In terms of the canonical variables the fixed points in the two parameter regimes are
\begin{equation}\label{ss1}
\begin{array}{l}
(z,\phi)=\left(\pm\sqrt{1-\lambda^2},\pi/2\right),
\end{array}
\end{equation}
for $0\leq\lambda\leq1$, and 
\begin{equation}\label{ss2}
\begin{array}{l}
(z,\phi)=\left(\pi/2,\pm\sqrt{1-\lambda^{-2}}\right),
\end{array}
\end{equation}
for $\lambda\geq1$, and with $\lambda=\omega/2\kappa$ defined in the main text. 
 
The Jacobians corresponding to solutions~(\ref{ss1}) are given by
\begin{equation}
J=\left[
\begin{array}{cc}
\pm2\kappa\sqrt{1-\lambda} & 0\\
0 & \pm2\kappa\sqrt{1-\lambda}
\end{array}\right]
\end{equation}
and it is clear that only the second solution gives negative eigenvalues and is thereby stable. Thus, while the solutions are symmetric under the reflection $S_z\leftrightarrow-S_z$, the results from the stability analysis show that the two solutions are qualitatively different which prohibits any such parity symmetry. The negative eigenvalues, here both identical, give the time-scale $T=1/2\kappa\sqrt{1-\lambda}$ for relaxing to the steady state (in the validity regime for the linear expansion). Note that provided the mean-field analysis gives an accurate description in the thermodynamic limit in the magnetized phase, the above result tells us that the gap $\Delta$ of the Liouvillian $\hat{\mathcal L}$ closes as $\Delta\propto\kappa\sqrt{1-\lambda}$. 

The remaining two solutions, Eq.~(\ref{ss2}), give the Jacobians
\begin{equation}
J=\left[
\begin{array}{cc}
0 & \pm\omega\sqrt{1-\lambda^{-1}} \\
\mp\omega\sqrt{1-\lambda^{-1}} & 0
\end{array}\right]
\end{equation}
with purely imaginary eigenvalues. This is reminiscent of a Hopf bifurcation, but contrary to a Hopf here the trajectories do not approach limit cycles~\cite{strogatz}. The appearance of periodic solutions for $\lambda>1$ was already predicted in Ref.~\cite{model1sm} using a Fokker-Planck method for the Glauber distribution. But the actual steady states were not identified in that reference.
   
 \begin{figure}
\centerline{\includegraphics[width=6cm]{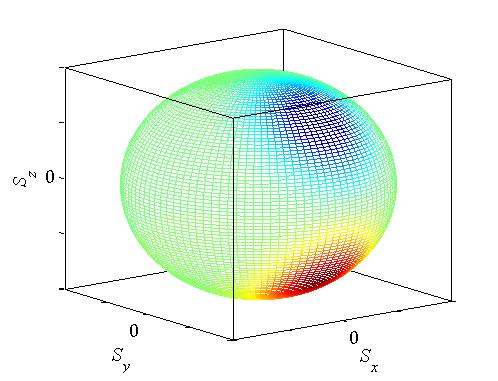}}
\caption{(Color online) A schematic figure demonstrating the idea of one `repulsive' (unstable) one `attractive' (stable) fixed point on the phase space of the spin. This plot corresponds to a coupling $\lambda=0.35$ where the fixed points are somewhere between the north (south) and east poles. The system approaches the stable fixed point in a finite time regardless of its initial state. On a flat phase space there would be a small set of initial states that would not reach the stable fixed point in a finite time. } 
\label{fig1}
\end{figure}  

The equations for $z$ and $\phi$ cannot be put on a `potential form', i.e. there exist no local function $V(z,\phi)$ such that $\frac{\partial z}{\partial t}=\frac{\partial V(z,\phi)}{\partial\phi}$ and $\frac{\partial\phi}{\partial t}=-\frac{\partial V(z,\phi)}{\partial z}$. This, of course, derives from the dissipative nature of the problem. We may, however, schematically think of the fixed points as attractors or repellers on the phase space. Since the phase space is a sphere, in the magnetized phase we should envision one repulsive and one attractive point on the sphere along the $yz$-meridian, as illustrated in Fig.~\ref{fig1}. Numerically we have verified that for random initial states, in a rather short time the state approaches the stable fixed point. This fast relaxation of general states (possibly far from the stable fixed point) is believed to be achievable due to the geometry of the phase space. On a plane, with one attractive and one repulsive fixed point one would most likely find states that relax infinitely slow. Indeed, the phase space geometry, together with the fact that the model is dissipative, might also explain why the bifurcations are not any of the standard ones. This, in turn, is crucial for finding a continuous bifurcation without reflecting any underlying symmetry.  
 
\section{Absence of symmetry breaking}
The symmetries of a closed system are embedded in the eigenstates of its Hamiltonian. For a model supporting a $\mathbb{Z}_2$ symmetry for example, the ground state is an even parity state. In the thermodynamic limit the first excited state (with odd parity) becomes degenerate with the ground state in the symmetry broken phase. The symmetries are naturally also reflected in the phase space distributions. In particular for the $\mathbb{Z}_2$ symmetry one finds for the ground state two blobs with interferences in between. Thus, by mapping out the phase space distributions as a function of $\lambda$ one can identify the phase transition and visualize its accompanying symmetry breaking. In NESS PT's, where a symmetry breaking occurs, one typically encounters the same qualitative behavior as for QPTs~\cite{joshi}.

\begin{figure}
\centerline{\includegraphics[width=8cm]{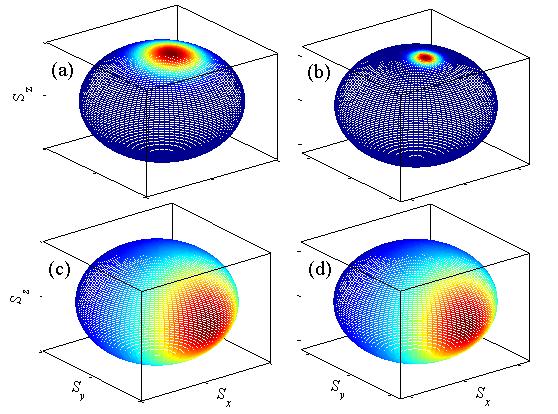}}
\caption{(Color online) The Husimi $Q$-function for $\lambda=0.05$, (a) and (b), and $\lambda=1.05$, $(c)$ and $(d)$. For the left panel $S=10$ and $S=100$ for the right panel. In the first two plots, in the magnitized phase, the state is approximately pure and the fluctuations are quantum. This explains the decreasing amount of fluctuations for increasing $S$; in the thermodynamic (classical) limit $S\rightarrow\infty$ the distribution would collapse to a single point on the phase space (the Planck-cell becomes vanishingly small in comparison to the whole phase space). The last two plots depict the distributions in the incoherent phase and evidently the amount of fluctuations is considerably larger. These stem from the openness of the model, and in particular as $\lambda\rightarrow\infty$ or $S\rightarrow\infty$ the state is maximally mixed and the corresponding $Q$-function will be smeared out entirely over the sphere. The $z$-axis has been flipped in order to better visualize the distribution.} 
\label{fig2}
\end{figure}

Drawing from our conclusions, the situation should be different in the present model; one should only find a single blob in the phase space distribution. However, once crossing the critical point from the magnetized phase, fluctuations should greatly set in and start to smear out the distribution. As a demonstration we consider the $SU(2)$ $Q$-function defined as~\cite{spinq}
\begin{equation}
Q(z)=\langle z|\hat\rho|z\rangle,
\end{equation}
with the spin coherent state~\cite{spinc}   
\begin{equation}
|z\rangle=\frac{1}{\left(1+|z|^2\right)^j}\sum_{m=-S}^{S}\sqrt{\left(\begin{array}{c}
K\\
S+m\end{array}\right)}z^{S+m}|S,m\rangle,
\end{equation}
where $z=\mathrm{e}^{i\phi}\tan\frac{\theta}{2}$, $|S,m\rangle$ is the spin angular momentum state with $z$-quantum number $m$, and where $K$, as in the main text, is the number of qubits. The results for the $Q$-function for two different spins are shown in Fig.~\ref{fig2} (see also Ref.~\cite{model4sm}). We indeed see no indications of a symmetry breaking, and it confirms the picture of fluctuations blowing up in the incoherent phase. In particular, to a good agreement the size of $S$ determines only the amount of fluctuations in the magnetized phase where we know that in the classical limit $S\rightarrow\infty$ the distribution collapses into a single point.

\begin{figure}
\centerline{\includegraphics[width=8cm]{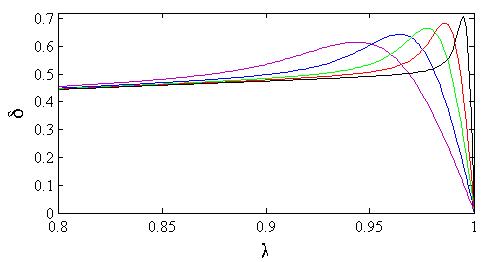}}
\caption{(Color online) The numerically calculated critical exponent for the magnetization $S_z$ for different system sizes; $S=50$ (magenta), 100 (blue), 200 (green), 4000 (red), and 1600 (black). As the critical point $\lambda_c=1$ is approached one easily convinces oneself that $\delta=1/2$ in the thermodynamic limit, which is in agreement with the mean-field result.} 
\label{fig3}
\end{figure}

\section{Quantum critical exponents}
For a continuous PT we expect that sufficiently close to the critical point any local observable
\begin{equation}
\langle\hat{\mathcal O}\rangle\propto|\lambda-\lambda_c|^{\beta_\pm},
\end{equation}
where $\beta_\pm$ is the exponent depending on whether the critical point is approached from above or below, i.e. the behaviour need not be symmetric. By local in our model we mean that $\hat{\mathcal O}=\sum_i\hat{o}_i$ with $\hat{o}_i$ a single particle operator acting on qubit $i$. In the main text it was mentioned that for the magnetization $\hat S_z$ (which is clearly a local operator) the critical exponent $\delta=1/2$ both at the mean-field and full quantum level. The numerical results for the critical exponent, confirming that $\delta=1/2$, are presented in Fig.~\ref{fig3}. In more general terms, for operators $\hat{\mathcal C}=\sum_i\sum_j\cdots\sum_k\hat a_i\hat b_j\cdots\hat c_k$ expanded in single particle operators we define `locality' by the number of single particle operators in the product $\hat a_i\hat b_j\cdots\hat c_k$. 

\begin{figure}
\centerline{\includegraphics[width=8cm]{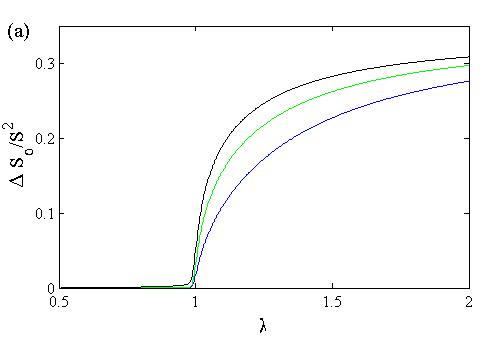}}
\centerline{\includegraphics[width=8cm]{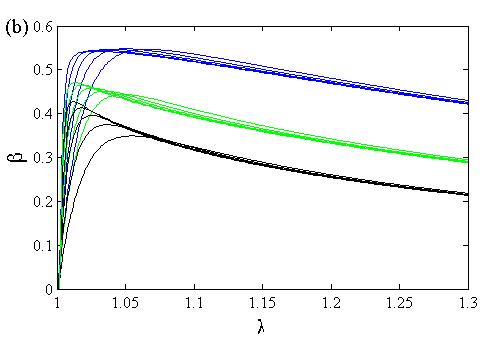}}
\caption{(Color online) The variances $\Delta S_\alpha$ ($\alpha=x,\,y,\,z$, blue, green, and black respectively) (a) and their respective exponents (b) as a function of $\lambda$. The different curves in (b) are the results for $S=50,\,100,\,200,\,400,\,1600$, and in (a) $S=400$. By extrapolating the results of (b) to $S=\infty$ one finds $\beta_x\approx0.54$, $\beta_y\approx0.48$, and $\beta_z\approx0.46$ at the critical point.} 
\label{fig4}
\end{figure}

Within this definition, the spin variances
\begin{equation}
\Delta S_\alpha=\langle\hat S_\alpha^2\rangle-\langle\hat S_\alpha\rangle^2,\hspace{1cm}\alpha=x,\,y,\,z
\end{equation}
are examples of operators with a weak non-local contribution, only containing products of single particle operators. The three variances for different $S$ are displayed in Fig.~\ref{fig4} (a), and the corresponding exponents in (b). By extrapolating the results to $S=\infty$ the three exponents seem to attain different values and in particular not simple fractional values (see figure caption). Normally for fully connected models the mean-field results are correct and one therefore expects simple fractions for the exponents~\cite{full}. In the present model we have seen, however, that in the incoherent phase the mean-field results do not agree with the full quantum ones. Note further that to derive the mean-field exponents for the variances one would need to go beyond the simple factorization of operators to include products of operators.

\begin{figure}  
\centerline{\includegraphics[width=8cm]{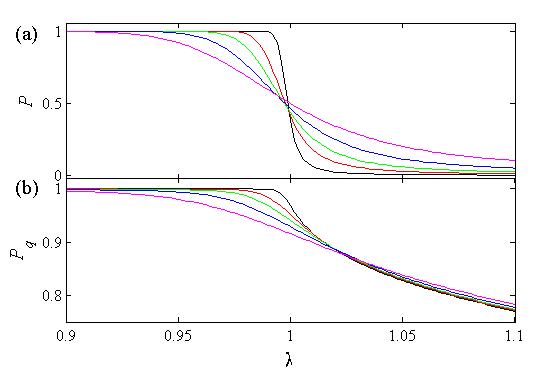}}  
\centerline{\includegraphics[width=8cm]{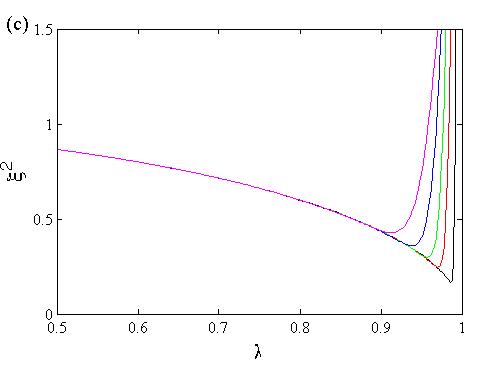}} 
\caption{(Color online) Purities for the full state $\hat\rho_\mathrm{ss}$ (a) and the single qubit reduced state $\hat\rho_\mathrm{q}$ (b), and full state $x$-squeezing (c) for the same spin sizes $S$ as in Fig.~\ref{fig3}. In the the thermodynamic limit $S\rightarrow\infty$, the purity of the state $\hat\rho_\mathrm{ss}$ shows a step like behaviour going discontinuously from a pure to a maximally mixed state. In the same limit the reduced single particle state stays continues with, however, a cusp-like behaviour. The squeezing is maximum at the critical point $\lambda=1$ as $S\rightarrow\infty$, and all squeezing is lost beyond the critical point $\lambda>1$. } 
\label{fig5}
\end{figure}

In Fig.~\ref{fig5} (a) the purity~\cite{pur}
\begin{equation}
P=\mathrm{Tr}\left[\hat\rho_\mathrm{ss}^2\right]
\end{equation}
is shown. The purity is a measure of how mixed the state is, with $P=1$ representing a pure state and $P=1/(2S+1)$ representing the fully mixed state (i.e. the density operator proportional to the identity matrix). We may alternatively interpret the purity as the expectation value of the state itself, $P=\langle\hat\rho_\mathrm{ss}\rangle$. From the explicit form of $\hat\rho_\mathrm{ss}$ (Eq.~(3) in the main text) this is an example of a maximally delocalized observable. As we see in the Fig.~\ref{fig5}, when $S\rightarrow\infty$ $P$ becomes discontinuous and jumps from $P=1$ in the magnetized phase to $P=0$ in the incoherent phase. This is thereby an example of a quantity that is not continuous across the critical point and one may object that the transition should not be viewed as a continuous one. However, one must remember that $P$ is the expectation value of a maximally delocalized operator, and one can find similar discontinuities in equilibrium continuous QPT's. An example is the fidelity susceptibility~\cite{fid}. As a simplified version of the susceptibility consider the operator $\hat\Psi_\varepsilon=|\psi_0(\lambda+\varepsilon)\rangle\langle\psi_0(\lambda+\varepsilon)|$, where $|\psi_0(\lambda)\rangle$ is the ground state for the coupling $\lambda$ and $\varepsilon$ some small number. The expectation of this operator, $\chi_\varepsilon=\langle\psi_0(\lambda)|\hat\Psi_\varepsilon|\psi_0(\lambda)\rangle$, will be discontinuous as $\lambda$ approaches $\lambda_c$. If we would instead consider the purity $P_\mathrm{q}$ for a single qubit this is a local operator and we should not encounter a discontinuity. This is confirmed in Fig.~\ref{fig5} (b), where we in particular see the characteristic cusp-like behavior for a continuous PT as $S\rightarrow\infty$. This purity is in this case defined equivalently; $P_\mathrm{q}=\mathrm{Tr}\left[\hat\rho_\mathrm{q}^2\right]$ with $\hat\rho_\mathrm{q}$ the reduced density operator for a single qubit. A similar behavior (not shown here) is also found for the purity for the two-qubit state of Eq.~(\ref{2q}), but with a slightly `sharper' emerging discontinuity.

\section{Quantum properties of $\hat\rho_\mathrm{ss}$}
Properties of quantum correlations in the present model have been discussed in Refs.~\cite{model4,model3sm}. In Fig. 2 of the main text we showed how the negativity of a two-qubit state behaves with system size and the coupling parameter $\lambda$. The negativity is given by~\cite{negsm} 
\begin{equation}
\mathcal{N}=\sum_i\frac{|\mu_i|-\mu_i}{2},
\end{equation}
where $\mu_i$ is the $i$'th eigenvalue of the partially transposed reduced density operator $\hat\rho_\mathrm{2q}^{T_A}$ for the two qubits. Whenever the full multi-qubit state $\hat\rho$ is symmetric with respect to exchanging qubits, the two-qubit reduced density operator is easily evaluated~\cite{molmersm}. In particular, the reduced state becomes
\begin{equation}\label{2q}
\hat\rho_\mathrm{2q}=\left[
\begin{array}{cccc}
v_+ & x_+^* & x_+^* & u^*\\
x_+ & w & w & x_-^*\\
x_+ & w & w & x_-^*\\
u & x_- & x_- & v_-
\end{array}\right],
\end{equation}
with the elements expressed in the collective spin expectations 
\begin{equation}
\begin{array}{l}
\displaystyle{v_\pm=\frac{K^2-2K+4\langle\hat S_z^2\rangle\pm4\langle\hat S_z\rangle(K-1)}{4K(K-1)}},\\ \\
\displaystyle{x_\pm=\frac{(K-1)\langle\hat S_+\rangle\pm\langle[\hat S_+,\hat S_z]_+\rangle}{2K(K-1)}},\\ \\
\displaystyle{w=\frac{K^2-4\langle\hat S_z^2\rangle}{4K(K-1)}},\\ \\
\displaystyle{u=\frac{\langle\hat S_+^2\rangle}{K(K-1)}}.
\end{array}
\end{equation}
From Eq.~(\ref{2q}) one also directly obtains the single qubit reduced density operator
\begin{equation}
\hat\rho_\mathrm{q}=\left[
\begin{array}{cc}
w+v_+ & x_+^*+x_-^*\\
x_++x_- & w+v_-\end{array}\right].
\end{equation} 

From Fig.~\ref{fig5} (b) it is seen that in the thermodynamic limit, the reduced density operators are pure in the magnetized phase, and continuously become more and more mixed in the incoherent phase. At the same time we saw in Fig. 2 of the main text that qubit-qubit entanglement vanishes in the incoherent phase for any $S$. The fact that $P_\mathrm{q}=1$ in the thermodynamic limit in the magnetized phase says that there cannot be any quantum correlations surviving as $S\rightarrow\infty$. To explore this further we consider the spin squeezing~\cite{squeez} 
\begin{equation}
\xi^2=\frac{2S\Delta S_{n_1}^2}{\langle\hat S_{n_2}\rangle^2+\langle\hat S_{n_3}\rangle^2}.
\end{equation}
Here, $n_1$, $n_2$, and $n_3$ are three mutually orthogonal vectors, and we restrict ourselves to $x$, $y$, and $z$. Whenever $\xi^2<1$ the state is squeezed, and in addition, spin squeezing also acts as an entanglement witness~\cite{squeez,ew}. This means that if $\xi^2<1$ the state cannot be cast on a product form, and hence it must embody some sort of quantum correlations. For $n_1=y$ or $z$ one finds that $\xi^2>1$ for any $S$ and $\lambda$. However, as demonstrated in Fig.~\ref{fig5} (c), for $n_1=x$ the state is squeezed for the majority of couplings $\lambda<1$ and all $S$, and hence, the state is indeed entangled. In the thermodynamic limit maximum squeezing is obtained at the critical point (presumably $\xi^2\rightarrow0$ when $S\rightarrow\infty$), and for smaller couplings $\lambda$ the dependence on $S$ is very weak meaning that the amount of multipartite-correlations in the state remains when $S$ is increased. At first this seems to contradict the results of Fig.~\ref{fig5} (b). However, it must be remembered that $\xi^2$ says something about the total amount of quantum correlations, while the purity is for single qubits, and thus even though the state factorizes in the limit $S\rightarrow\infty$ as long as $S$ is finite there is a small mixture and it comprises the finite full state entanglement. This is the idea behind entanglement sharing~\cite{esharesm}. The shared entanglement is further motivated by noting (numerically) that the negative scales as $N\sim S^{-1}$ for $\lambda>1$ away from the critical point. Closer to the critical point, the $S$-dependence of the squeezing becomes evident and the scaling of the negativity is also weaker, $N\sim S^{-0.9}$.



\end{document}